\documentstyle[preprint,prl,aps]{revtex}
\begin{document}
\draft
\title{Saddle Points and Stark Ladders: Exact Calculations of Exciton Spectra in
Superlattices}
\author{D. M. Whittaker}
\address{Department of Physics, University of Sheffield, Sheffield. S3 7RH. UK.}
\date{August 24, 1993}
\maketitle
\begin{abstract}
A new, exact method for calculating excitonic absorption in superlattices is
described. It is used to obtain high resolution spectra showing the
saddle point exciton feature near the top of the miniband. The evolution of
this feature is followed through a series of structures with increasing
miniband width. The Stark ladder of peaks produced by an axial electric field
is investigated, and it is shown that for weak fields the line shapes are
strongly modified by coupling to continuum states, taking the form of Fano
resonances.  The calculated spectra, when suitably broadened, are found to be
in good agreement with experimental results.
\end{abstract}

A proper treatment of excitonic effects is essential to the understanding of the
optical properties of semiconductors. However, in superlattices it is difficult
to calculate exciton spectra because of the anisotropic, quasi-three dimensional
nature of the problem. In this letter, the first high resolution
theoretical superlattice absorption spectra are presented, calculated using a
new method which generalizes the Green's function approach of Zimmermann
\cite{zimmer} for quantum wells.  Particular attention is paid to two aspects of
the spectra which historically have been of considerable interest in the theory
of semiconductors: the first is the saddle point exciton \cite{saddle}
associated with the top of the miniband, the second is the Fano resonance line
shape \cite{fano} predicted for Stark ladder peaks when an electric field is
applied to the superlattice.

A saddle point is a critical point in the bandstructure at
which the sign of the mass is different along different directions. For
superlattices, the important region is around the M1 critical point at the top
of the miniband, where the axial dispersion has a negative effective mass,
while the in-plane mass is positive. The associated excitonic feature is not a
sharp peak, but a broad enhancement occurring on the low energy side of the
saddle point. There has been a long theoretical interest in the behavior of
this feature in bulk semiconductors \cite{saddle}, and some low
resolution calculations exist for superlattices \cite{ccc,my}. In this letter,
high resolution spectra are used to trace the development of the superlattice
saddle point exciton as the miniband width is increased. It is
shown that weak structure occurs in the enhancement, corresponding to
excited saddle point states.

A Stark ladder occurs as a result of the localization of miniband states when an
axial electric field is applied to a superlattice. The localized states are
peaked in individual wells and their energies form a ladder of equally spaced
levels, with separation equal to the potential drop across a period of the
structure. Transitions between different electron and hole states lead to a
series of two dimensional like steps in the absorption spectrum, each with an
associated exciton peak. For large fields, these are well defined and
adequately treated by previous theoretical methods
\cite{my,ds}. However, it is shown in this letter that,
at low fields, the mixing between the excitons and the surrounding continuum
states cannot be ignored and leads to Fano resonances
\cite{fano} with significant widths. The possibility of studying Fano
resonances in low dimensional structures has been the subject of much
theoretical discussion \cite{fanores}. There has, however, been no clear
experimental demonstration of the characteristic spectral line shapes, although
time resolved measurements show the existence of homogeneous broadening
due to resonant coupling \cite{coupling}. The present calculations suggest that
Fano line shapes should be observable in Stark ladder spectra from high quality
superlattices.

The problem of the exciton in a superlattice has been considered theoretically
by a number of authors, in treatments both with \cite{my,ds} and without
\cite{ccc,my,chomette,young} an electric field. A number of methods have
been used, with good results for the energies of the bound and quasi-bound
states. However, the only spectra which are obtained \cite{ccc,my,young} are
fairly crude, with the continuum features approximated by sets of discrete peaks
which need broadening to give realistic results. Though adequate for comparison
with most experimental spectra, the broadening makes it impossible to resolve
the detailed structure which is the subject of the present work.

The approach adopted here starts from an effective Hamiltonian which
simplifies the exciton problem without loss of accuracy. This is solved
numerically using a new method which avoids the problems with resolution.  The
spectra so obtained are essentially exact, and the calculation remains
computationally feasible for arbitrarily small broadening.

The effective Hamiltonian is derived by
restricting the states used in constructing the exciton to a single pair of
minibands (one each for electron and hole) and diagonalizing the Coulomb
interaction and electric field terms exactly within this basis \cite{my}.
This approximation is usually very good, because the miniband separations in
typical superlattices tend to be large compared to the energy scale of the
excitonic effects.

For the present purpose, the miniband states can best be described in a
representation based on localized Wannier functions \cite{wannier}. The
wavefunction for optically active excitons can then be written in the form
\begin{equation}
\Psi(z_e,z_h,r) = \sum_{n_e,n_h} \psi(n_e-n_h,r) f_e(z_e-n_ed) f_h(z_h-n_hd).
\end{equation}
Here, $\psi(n,r)$ is a superlattice scale envelope function which depends only
on the separations of the electron and hole: $n=n_e-n_h$ (the number of periods
of width $d$) along the growth axis and $r$ in the plane of the wells.
$f_e(z_e)$ and $f_h(z_h)$ are the Wannier functions for the pair of minibands
under consideration.

In the Wannier representation, the miniband kinetic energy takes the form
of a tight-binding type hopping operator:
\begin{equation}
T_{\text{hop}}\, \psi(n,r) = \sum_i \Delta_i \left[\psi(n+i,r) +
\psi(n-i,r)\right]
\end{equation}
where the hopping terms $\Delta_i$ are obtained from a Fourier
expansion of the miniband dispersion. The diagonal matrix element of the Coulomb
interaction can be written as an effective potential
\begin{equation}
V_{\text{eff}}(n,r) = - \frac{1}{\varepsilon_r} \int \! \! \int dz_e \, dz_h \,
\frac{ |f_e(z_e)|^2 |f_h(z_h)|^2} {\sqrt{(z_e-z_h+nd)^2 + r^2}}.
\end{equation}
There are also small off-diagonal matrix elements, but these are
always negligible in comparison with $V_{\text{eff}}$ and the superlattice
coupling.

The envelope function satisfies a Schr\"{o}dinger equation with Hamiltonian
\begin{equation}
\label{ham}
H_{\text{eff}} = - \frac{1}{2m} \nabla_r^2 + T_{\text{hop}} +
V_{\text{eff}}(n,r) + eFnd
\end{equation}
with $m$ the in-plane reduced mass and $F$ the electric field.
Eq.\ (\ref{ham}) is, despite its tight binding form, an essentially exact
formulation of the superlattice exciton within a single pair of minibands, and
is valid for minibands of any width.

Optical spectra are obtained from Eq.\ (\ref{ham}) using a generalization of
the method developed by Zimmermann \cite{zimmer} for quantum well excitons.
The absorption
coefficient is calculated by taking the limit $r,r' \rightarrow 0$ of
a Green's function $G(n,r;n',r';E)$ satisfying the inhomogeneous system
of equations:
\begin{equation}
\label{green}
[H_{\text{eff}}-E] \, G(n, r; n',r';E) = \delta_{n,n'} \, \delta (r-r')
\end{equation}
The Green's function is calculated from numerical solutions to Eq.\
(\ref{green}) which satisfy the boundary conditions at $r \rightarrow 0$ or
$r \rightarrow \infty$, using a straight forward generalization of the standard
approach for the two point boundary value problem \cite{green}.

Figs.\ \ref{fig1},\ref{fig2} show the calculated zero field absorption
spectra for a series of superlattices with miniband widths from 0 to 57 meV. The
structures considered all have GaAs wells of
width $L_w$=30 \AA\ and Al$_{0.35}$Ga$_{0.65}$As barriers with  widths $L_b$
ranging from 150 \AA\ to 30 \AA.

In Fig.\ \ref{fig1} results are shown for the wider barrier structures, where
the minibands are narrow, with widths smaller than, or comparable to, the
exciton binding energy. As $L_b$ is reduced and the miniband width increases,
the most obvious change is a reduction in the binding energy and oscillator
strength of the main exciton peak. This corresponds to the ground state
wavefunction spreading in the axial direction, becoming more three dimensional
in form \cite{chomette}. Changes also occur in the higher exciton states. In the
two dimensional limit, these are in-plane $s$ states ($2s,3s \ldots$), but for
finite miniband widths, an additional, more tightly bound state appears (labeled
$a$ in Fig.\ \ref{fig1}), becoming progressively stronger as the width
increases. Though it is not possible to describe this state with exact
 quantum numbers, its wavefunction contains large contributions from
in-plane 1$s$ states with electron and hole in adjacent periods, mixed in by the
superlattice coupling \cite{my}.

The narrow minibands do not have enough width to support the main
saddle point feature. There is, however, some excitonic structure apparent in
the continuum. This consists of a series of oscillations just below the M1
critical point, which correspond to the higher states of the saddle point
exciton. These oscillations broaden and weaken as the miniband becomes wider
and the underlying density of states increases.  Some evidence of this type of
behavior can been seen in lower resolution calculations \cite{ccc,my}, but the
oscillations seem to be too weak to be experimentally observable in presently
available structures.  Predictions of similar oscillations in bulk
semiconductors have been made by Baslev \cite{saddle}.

Fig.\ \ref{fig2} corresponds to the narrower barrier structures,
where the minibands are wide, with widths larger than the exciton binding
energy.  As the miniband width increases, the main saddle point exciton feature
develops: the shape of the continuum changes from dropping away at the band edge
to rising gradually towards a broad peak some way below the M1 critical point.
For wider minibands, the overall strength of the saddle point feature becomes
greater, at the expense of the bound state, but its increasing width means that
the height of the enhancement decreases. The oscillations associated with the
higher states remain, but they too become broader.

Figs.\ \ref{fig3},\ref{fig4} show Stark ladder spectra for a
Ga$_{0.47}$In$_{0.53}$As-Al$_{0.24}$Ga$_{0.24}$In$_{0.52}$As superlattice with
$L_w$=39\AA\ and $L_b$=46\AA. This structure was chosen because of the
availability of experimental absorption spectra \cite{vois}, which
provide a more direct basis for theoretical comparison than the more usual
photocurrent spectra. A miniband width of 30 meV was used in the calculations,
though this is considerably lower than the value of $\sim$40 meV obtained from
the Kronig-Penney model \cite{vois}.

In Fig.\ \ref{fig3}, both experimental and broadened theoretical spectra are
shown. The theoretical spectra generally reproduce the experimental results
very well, especially on the low energy side. The additional experimental
absorption at higher energies is mainly due to the light hole transitions, which
are not included in the present theory. The poor agreement at $F$=7kV/cm is
probably a result of the electric field in the experiments being under estimated
at low biases, due to incomplete depletion. A much better fit can be obtained by
increasing the field in the calculation to $\sim$11kV/cm.

At zero field, the band edge and saddle point exciton
features are apparent, while at high field, the nearly fully localized two
dimensional exciton (labeled $n=0$) has the correct position and
strength. In between, the transitions to adjacent periods ($n=\pm 1$)
can be seen moving to higher and lower energy ($\sim \pm eFd$). This behavior
has been discussed by a number of authors \cite{my,ds} and further details
can be found in the references.

Fig.\ \ref{fig4} shows theoretical high resolution spectra for the same
structure. The zero field spectrum is similar to those of Fig.\ \ref{fig2},
though it should be noted that the excitonic features are weaker at a comparable
miniband width (see $L_b$ = 40\AA\ in Fig.\ \ref{fig2}), a result of the
lower electron mass in GaInAs. For finite fields a considerable amount of
structure is apparent in the spectra. At  $F$=14 kV/cm, the $n=0,\pm 1,-2$
excitons are clearly distinguishable, along with peaks due to the higher
in-plane states at the same axial separation. The main peaks all have a finite
width, which at $\sim$1 meV is much larger than the numerical broadening. The
line shapes of these peaks are characteristic of Fano resonances \cite{fano},
rising up above the continuum on one side, but falling well below it, almost to
zero, on the other.

The broadened Fano line shapes are a result of the coupling, by the Coulomb
interaction, between the main exciton peaks and resonant continuum states from
lower energy transitions. This is a purely excitonic effect, which occurs
because the Coulomb term allows tunneling between states of different in-plane
momentum. It should be distinguished from the lifetime broadening of Stark
localized states due to Zener tunneling between minibands \cite{krieger}, which
is spectrally insignificant in the field range of interest here, contributing
only $\sim 10^{-6}$ meV to the line widths at $F=28$ kV/cm.

At fields below 14kV/cm, the peaks move closer together and the Fano
line widths become larger, since the coupling is strongest when the wave-vectors
of the resonant continuum states are small. It becomes less appropriate to think
in terms of two dimensional excitons associated with localized Stark ladder
states; it is better to consider the applied electric field as perturbing the
exciton spectrum, giving it a periodic structure with period $\sim eFd$, but
with an envelope very similar to the zero field shape. At $F$=7 kV/cm, the
structure still takes the form of recognizable Fano resonance line shapes, but
with all the strength in the region of the band edge and saddle point exciton
features. When $F$ is reduced to 3.5 kV/cm, the width of the resonances becomes
large compared to their mutual separation, and they interfere. The resulting
spectrum, though basically periodic, contains very complicated structure with
unusual line shapes.  Calculations at lower field show that the structure
becomes finer in scale, but remains present even for $F<1$ kV/cm.

To conclude, a new method for solving coupled exciton problems has been
described, and applied to obtain very high resolution spectra for saddle point
and Stark ladder features. This has introduced new physics, in the form of
additional structure in the spectra and novel peak line shapes.  Though the
weaker structure is probably experimentally inaccessible, it seems likely that
small improvements in the quality of presently available superlattices would
allow the observation of excited saddle point exciton states and the Fano
line shapes of Stark ladder peaks.

\begin{figure}
\caption{Exciton spectra for GaAs-AlGaAs superlattices with narrow minibands,
of widths less than or comparable to the exciton binding energy. For each
structure, three spectra shown with different magnifications and numerical
broadening of 2 meV (lower), 1 meV (middle) and 0.1 meV (upper). M0 and M1
are the critical points at the top and bottom of the miniband.}
\label{fig1}
\end{figure}

\begin{figure}
\caption{Exciton spectra for GaAs-AlGaAs superlattices with wide minibands. For
each structure, two spectra are shown with different magnifications and
broadening of 2 meV (lower) and 0.2 meV (upper). The emergence of the broad
saddle point exciton feature (SPE) below the top of the miniband (M1) is
clear}
\label{fig2}
\end{figure}

\begin{figure}
\caption{Comparison of theoretical (solid line) results with  experimental
(dashed line) Stark ladder spectra for the AlGaInAs-GaInAs structure of Bleuse
\protect{\em et al} \protect\cite{vois}. The theoretical results have been
 broadened by 10 meV. The same normalizing factor is used at all
fields.}
\label{fig3}
\end{figure}

\begin{figure}
\caption{High resolution results corresponding to the low field spectra of Fig.\
\protect\ref{fig3}. The line width is 0.2meV, except for the dashed part of the
zero field spectrum which has been broadened to 2meV.}
\label{fig4}
\end{figure}


\begin{references}

\bibitem{zimmer} R. Zimmermann, Phys. Stat. Sol. (b) {\bf 135}, 681 (1986).

\bibitem{saddle}  See, for example, B. Velick\'{y} and J. Sak, Phys. Stat. Sol.
{\bf 16}, 147 (1966);  E. O. Kane, Phys. Rev. {\bf 180}, 852 (1969); I. Baslev,
Solid State Commun. {\bf 52} 351 (1984).

\bibitem{fano} U. Fano, Phys. Rev. {\bf 124}, 1866 (1961).

\bibitem{ccc} Hanyou Chu and Yia-Chung Chang, Phys. Rev. B {\bf 36}, 2946
(1987); Phys. Rev. B {\bf 39}, 10861 (1989).

\bibitem{my} D. M. Whittaker, Phys. Rev. B {\bf 41}, 3238 (1990); Superlattices
and Microstructures {\bf 7}, 375 (1990).

\bibitem{ds} M. M. Dignam and J. E. Sipe, Phys. Rev. Lett. {\bf 64}, 1797
(1990); Phys. Rev. B {\bf 43}, 4097 (1991).

\bibitem{fanores} See, for example, D. A. Broido and L. J. Sham, Phys. Rev. B
{\bf 34}, 3917 (1986); A. Pasquarello and L. C. Andreani, Phys. Rev. B {\bf 44},
3162 (1991).

\bibitem{coupling} L. Schultheis, A. Honold, J. Kuhl, K. K\"{o}hler and C. W.
Tu, Phys. Rev. B {\bf 34}, 9027 (1986).

\bibitem{chomette} A. Chomette, B. Lambert, B. Deveaud, F. Clerot, A. Regreny
and G. Bastard, Europhys. Lett. {\bf 4}, 461 (1987).

\bibitem{young} P. M. Young, P. M. Hui and H. Ehrenreich, Phys. Rev. B {\bf 44}
12969 (1991).

\bibitem{wannier} G. H. Wannier, Phys. Rev. {\bf 52}, 191 (1937).

\bibitem{green} See, for example, G. Barton, Elements of Green's Functions
and Propagation (Clarendon Press, 1989) pp51-54.

\bibitem{vois} J. Bleuse, P. Voisin, M. Allovon and M. Quillec, Appl. Phys.
Lett. {\bf 53}, 2632 (1988).

\bibitem{krieger} J. B. Krieger and G. J. Iafrate, Phys. Rev. B {\bf 35} 9644
(1987).

\end{references}
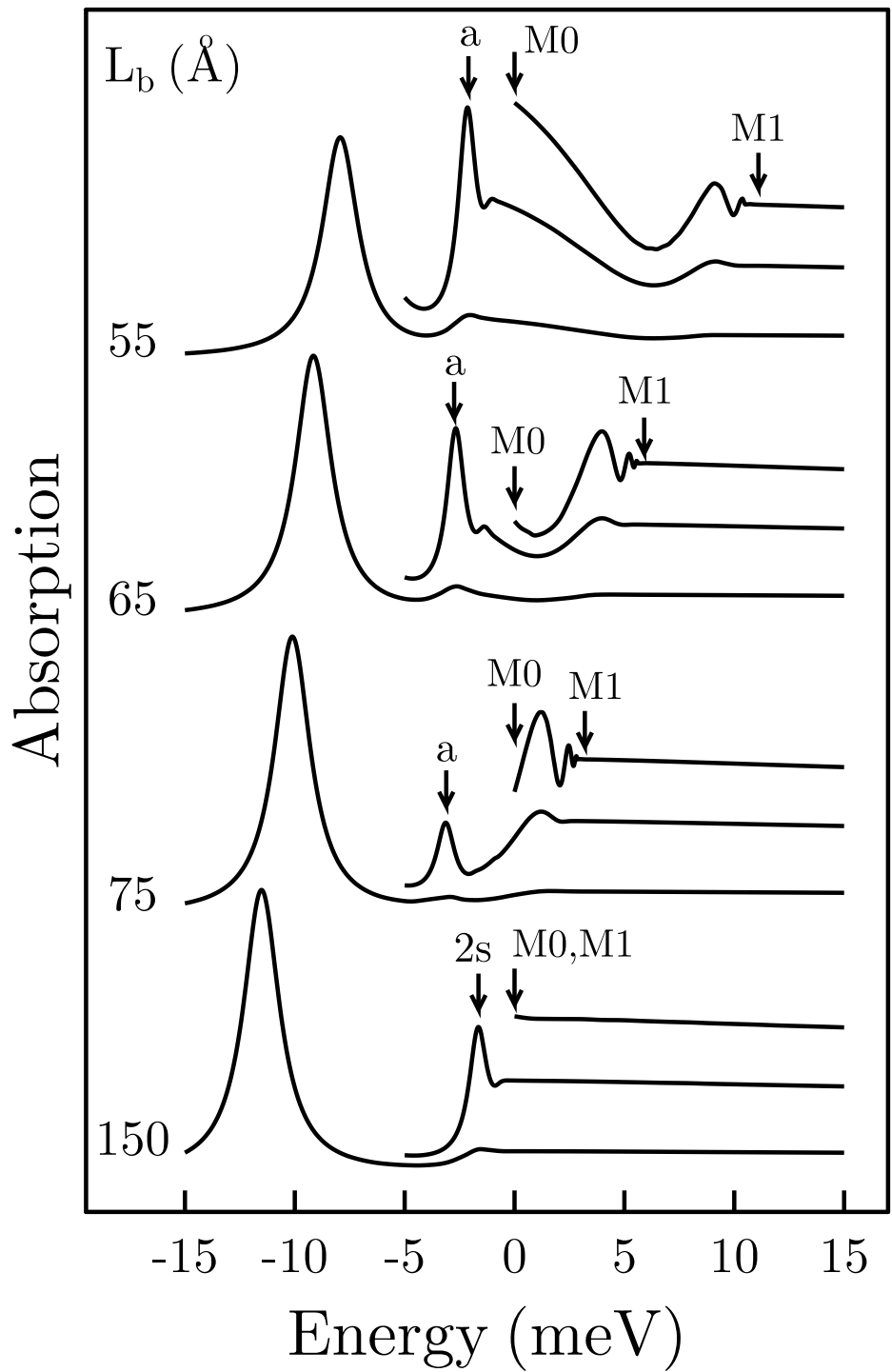
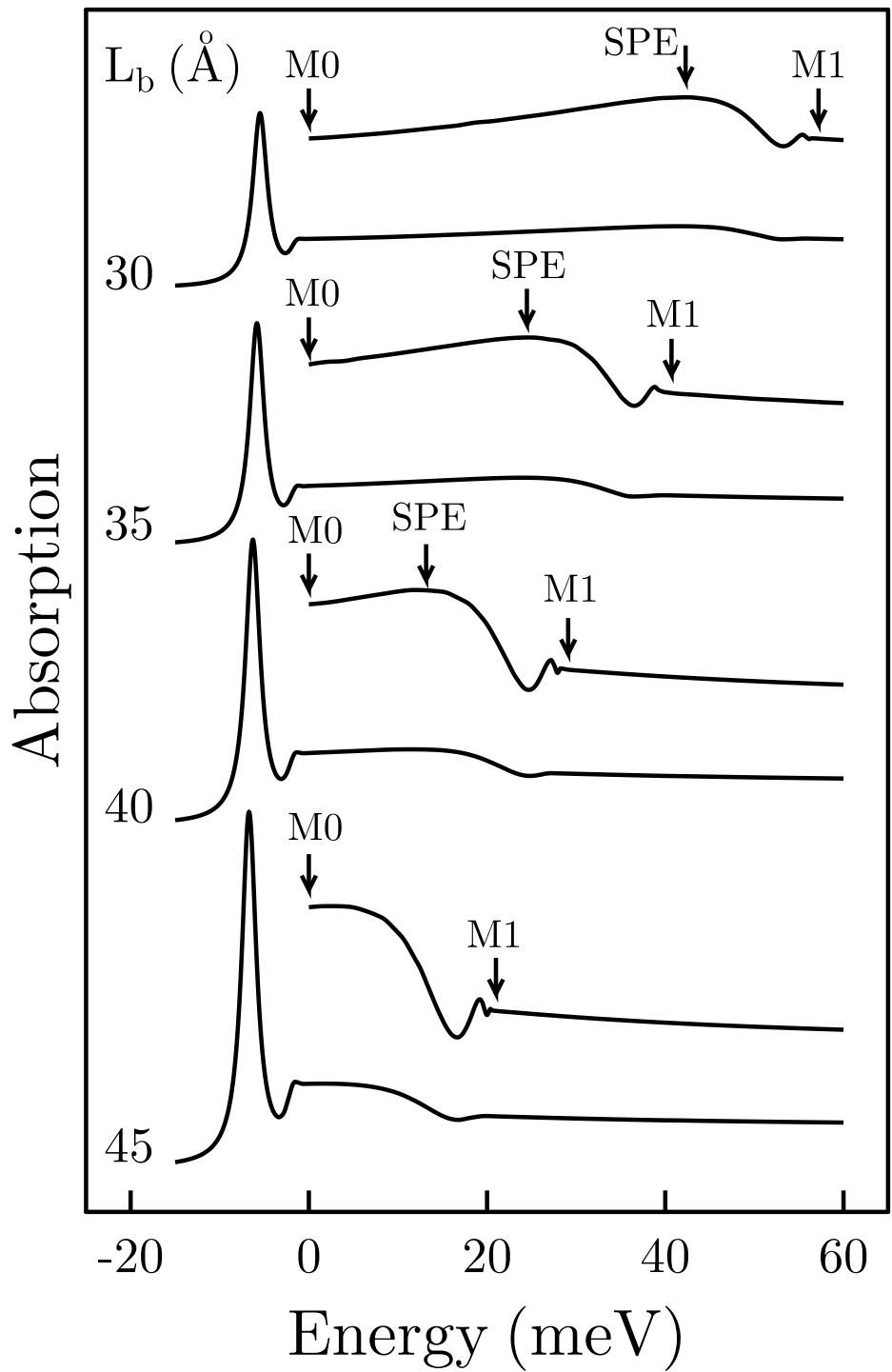
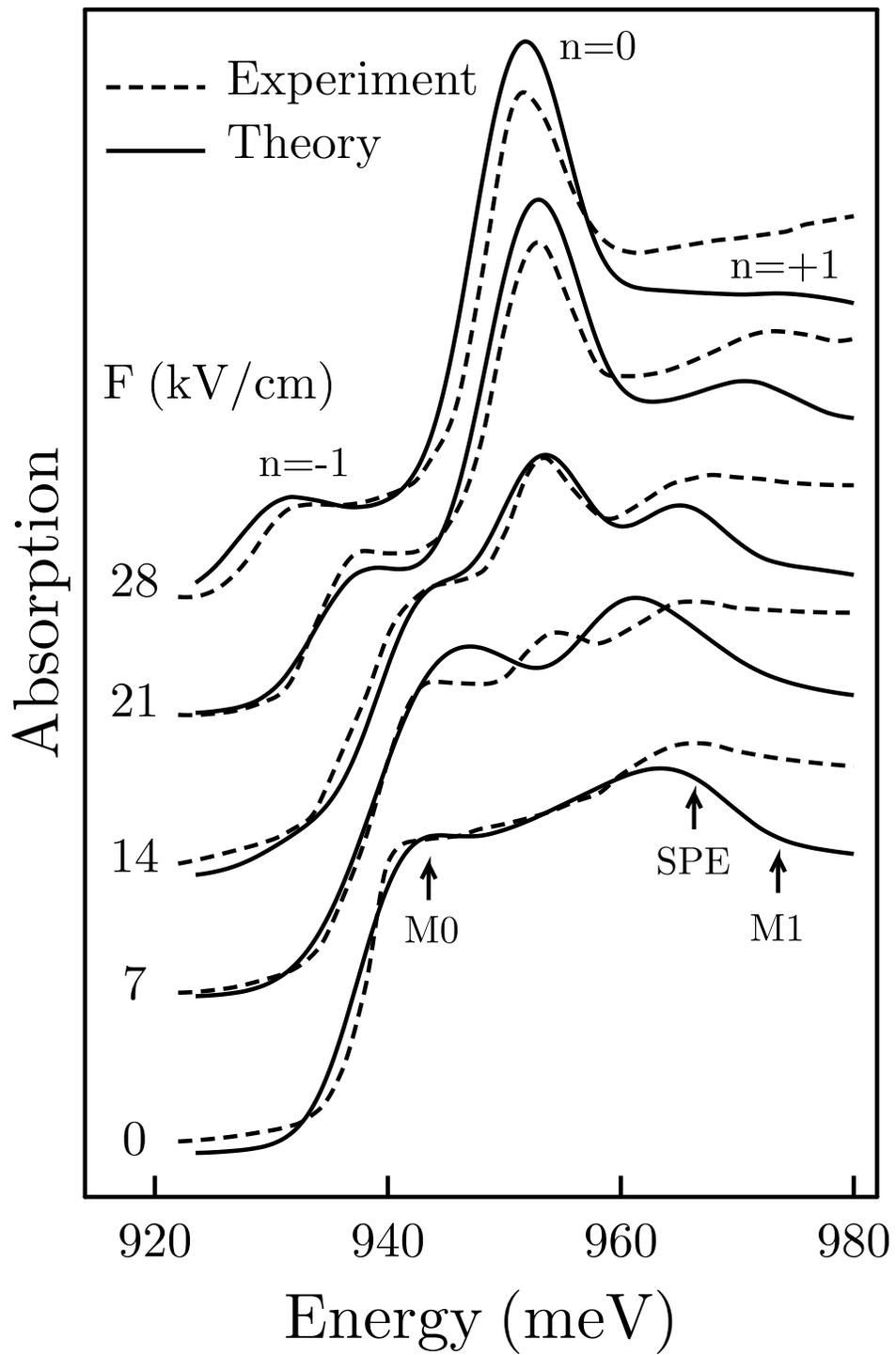
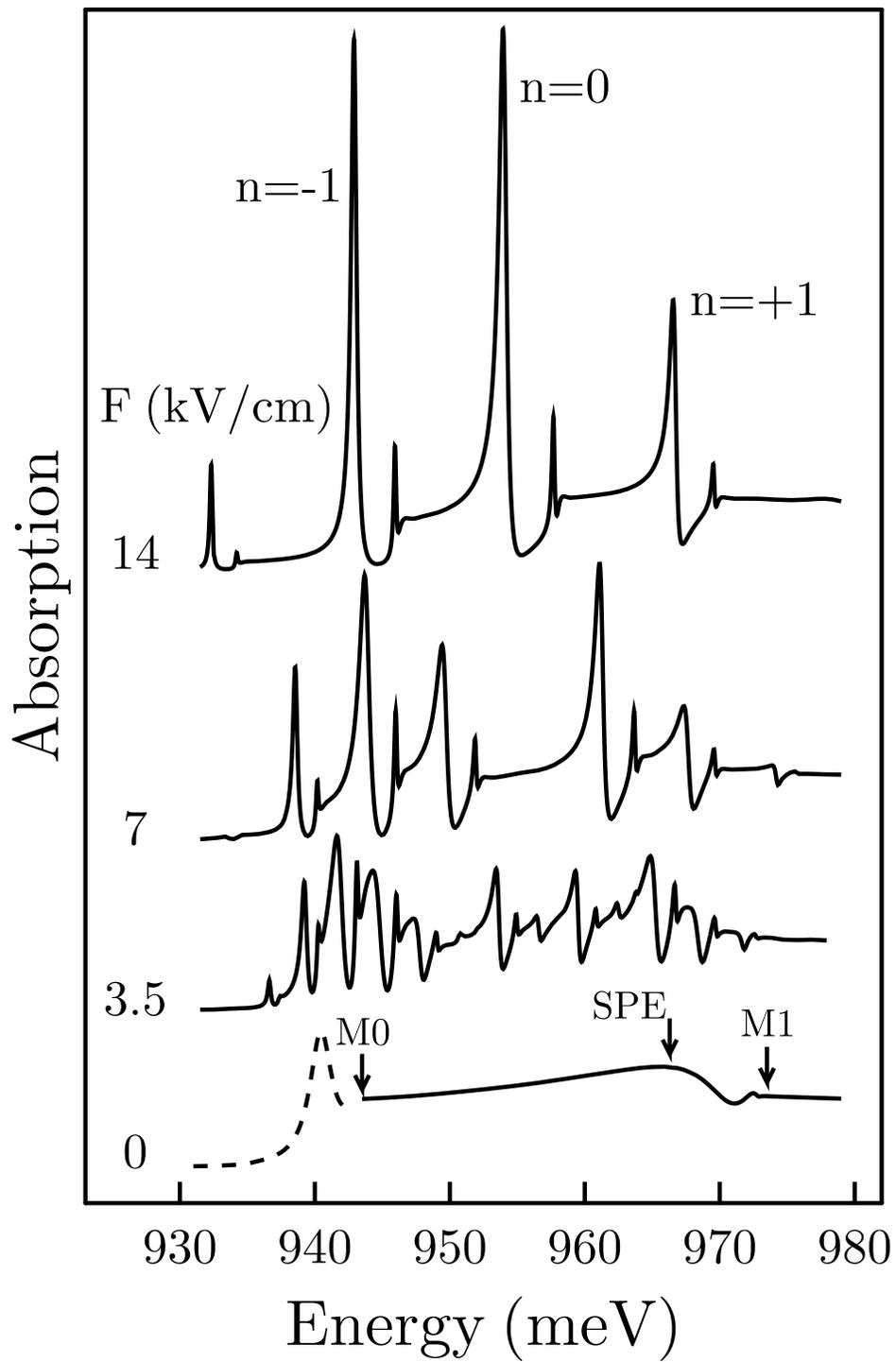
\end{document}